\documentclass{ifacconf}

\usepackage{graphicx}      
\usepackage{natbib}        
\usepackage{amsmath} 

\DeclareMathOperator*{\rrtstar}{\texttt{RRT}^{*}}
\usepackage{amssymb}  
\usepackage{algorithm}
\usepackage{subcaption}
\usepackage{todonotes}
\usepackage[noend]{algpseudocode}
\renewcommand{\algorithmiccomment}[1]{\State\bgroup//~#1\egroup}
\algnewcommand\algorithmicforeach{\textbf{for each}}
\algdef{S}[FOR]{ForEach}[1]{\algorithmicforeach\ #1\ \algorithmicdo}
\newcommand{\ts}{\textsuperscript}

\makeatletter
\let\old@ssect\@ssect 
\makeatother
\usepackage[colorlinks=true,allcolors=steelblue]{hyperref}
\makeatletter
\def\@ssect#1#2#3#4#5#6{%
  \NR@gettitle{#6}
  \old@ssect{#1}{#2}{#3}{#4}{#5}{#6}
}

\definecolor{steelblue}{RGB}{70,130,180}

\begin{document}
\begin{frontmatter}

\title{Towards Integrated Perception and Motion Planning with Distributionally Robust Risk Constraints\thanksref{footnoteinfo}} 

\thanks[footnoteinfo]{This work is partially supported by Defence Science and Technology Group, through agreement MyIP: ID9156 entitled ``Verifiable Hierarchical Sensing, Planning and Control'', the Australian Government, via grant AUSMURIB000001 associated with ONR MURI grant N00014-19-1-2571, and by the United States Air Force Office of Scientific Research under award number FA2386-19-1-4073.}

\author[First]{Venkatraman Renganathan} 
\author[Second]{Iman Shames} 
\author[First]{Tyler H. Summers}

\address[First]{Department of Mechanical Engineering, \\ University of Texas at Dallas, 
   Richardson, TX 75080 USA \\ (e-mails: \{vrengana, tyler.summers\}@utdallas.edu).}
\address[Second]{Department of Electrical and Electronic Engineering and Melbourne Information, Decision and Autonomous Systems (MIDAS) Laboratory, University of Melbourne, Parkville, VIC 3010, Australia \\ (e-mail: ishames@unimelb.edu.au)}

\begin{abstract}                
Safely deploying robots in uncertain and dynamic environments requires a systematic accounting of various risks, both within and across layers in an autonomy stack from perception to motion planning and control. Many widely used motion planning algorithms do not adequately incorporate inherent perception and prediction uncertainties, often ignoring them altogether or making questionable assumptions of Gaussianity. We propose a distributionally robust incremental sampling-based motion planning framework that explicitly and coherently incorporates perception and prediction uncertainties. We design output feedback policies and consider moment-based ambiguity sets of distributions to enforce probabilistic collision avoidance constraints under the worst-case distribution in the ambiguity set. Our solution approach, called Output Feedback Distributionally Robust $\rrtstar$(\texttt{OFDR}-$\rrtstar)$, produces asymptotically optimal risk-bounded trajectories for robots operating in dynamic, cluttered, and uncertain environments, explicitly incorporating mapping and localization error, stochastic process disturbances, unpredictable obstacle motion, and uncertain obstacle locations. Numerical experiments illustrate the effectiveness of the proposed algorithm.
\end{abstract}


\end{frontmatter}

\section{Introduction}
More sophisticated motion planning and control algorithms are needed for the robots to operate in increasingly dynamic and uncertain environments to ensure safe and effective autonomous behavior. Many widely used motion planning algorithms have been developed in deterministic settings. However, since motion planning algorithms must be coupled with the outputs of inherently uncertain perception systems, there is a crucial need for more tightly coupled perception and planning frameworks that explicitly incorporate perception uncertainties.

Motion planning under uncertainty has been considered in several lines of recent research \cite{blackmore_pioneer,agha_firm, luders_rrt, luders_rrtstar, blackmore_cc_mp, liu_risk_aware_mp,zhu2019chance}. Many approaches make questionable assumptions of Gaussianity and utilize chance constraints, ostensibly to maintain computational tractability. However, this can cause significant miscalculations of risk, and the underlying risk metrics do not necessarily possess desirable coherence properties \cite{rockafellar2007coherent, pavone_risk}. The emerging area of distributionally robust optimization (DRO) shows that stochastic uncertainty can be handled in much more sophisticated ways without sacrificing computational tractability \cite{dr_goh,wiesemann2014dro}. These approaches allow modelers to explicitly incorporate inherent ambiguity in probability distributions, rather than making overly strong structural assumptions on the distribution.

Traditionally, the perception and planning components in a robot autonomy stack are loosely coupled, in the sense that nominal estimates from the perception system may be used for planning, while inherent perception uncertainties are usually ignored. This paradigm is inherited in part from the classical separation of estimation and control in linear systems theory. However, in the presence of uncertainties and constraints, estimation and control should \emph{not} be separated; there are needs and opportunities to explicitly incorporate perception uncertainties into planning, both to mitigate risks of constraint violation \cite{blackmore_pioneer,florenceIPC,luders_rrt,summers_iros_2018,zhu2019chance} and to actively plan paths that improve perception \cite{davide_perception_plan}. 

\noindent \textit{Contributions:} In this paper, we take steps toward a tighter integration of perception and planning in autonomous robotic systems. Our main contributions are: 

\begin{itemize}
    \item We propose a distributionally robust incremental sampling-based motion planning framework that explicitly and coherently incorporates perception and prediction uncertainties. Our solution approach, called Output Feedback Distributionally Robust $\rrtstar$ (\texttt{OFDR}-$\rrtstar)$ (Algorithm 1), produces asymptotically optimal risk-bounded trajectories for robots operating in dynamic, cluttered, and uncertain environments, explicitly incorporating mapping and localization error, stochastic process disturbances, unpredictable obstacle motion, and uncertain obstacle locations. We design output feedback policies and consider moment-based ambiguity sets of distributions to enforce probabilistic collision avoidance constraints under the worst-case distribution in the ambiguity set (Algorithm 2).
    \item We demonstrate via numerical simulation results that it gives a more sophisticated and coherent risk quantification compared to an approach that accounts for uncertainty using Gaussian assumption, without increasing the computation complexity. 
\end{itemize} 
The rest of the paper is organized as follows. The dynamical model of the robot and the uncertainty modeling in the motion planning problem is discussed in section \ref{sec_robot_model}. Then, the proposed \texttt{OFDR}-$\rrtstar$ algorithm for motion planning is explained in section \ref{sec_dr_rrt}. The simulation results using a double integrator model are then presented in section \ref{sec_sim_results}. The paper is finally closed in section \ref{sec_conclusions} with a summary of results and with directions for the future research.
\section{Robot \& Environment Modeling} \label{sec_robot_model}
Consider a robot operating in an uncertain environment, $\mathcal{X} \subset \mathbb{R}^{n}$ cluttered with $n_0$ obstacles. We denote the set of obstacles as $\mathcal{B} = \{1,\dots,n_0\}$. The robot and the obstacles are modeled as a stochastic discrete-time linear system
\begin{align} 
    x_{t + 1} &= A x_t + B u_t + G w_{r,t}, \label{eqn_robot_dynamics}\\
    \mathcal{O}_{it} & = \mathcal{O}^{0}_{i} \oplus c_{it}, \quad i \in \mathcal{B}, \label{eqn_obs_dynamics} \\
    \mathcal{X}_{it} &= \Phi\left( \mathcal{O}_{it}\right), \quad i \in \mathcal{B}, \label{eqn_obstacle_dynamics}
\end{align}
where $x_t \in \mathbb{R}^n$ is the robot state at time $t, u_t \in \mathbb{R}^m$ is the input at time $t$, and $A$ and $B$ are the system dynamics matrix and input matrix, respectively. The process noise $w_{r,t} \in \mathbb{R}^n$ is a zero-mean random vector independent and identically distributed across time. The initial condition $x_0$ is subject to an uncertainty model, with the distribution of $x_0$ belonging to an ambiguity set, $P_{x_0} \in \mathcal{P}^{x}$. Moreover, $\mathcal{O}^{0}_{i} \subset \mathbb{R}^{n}$ represents the shape of obstacle $i$,  $c_{it} \in \mathbb{R}^{n}$ is a random vector that represents an uncertain obstacle location and motion, not necessarily zero-mean, with unknown distribution $P^{c}_{it} \in \mathcal{P}^{c}_{it}$, and $\oplus$ denotes set translation. The obstacles $\mathcal{O}_{it}, i \in \mathcal{B}$ are assumed to be convex polytopes. The state $\mathcal{X}_{it}$ of obstacle $i$ is defined by a set function $\Phi: 2^{\mathbb{R}^{n}} \rightarrow \mathbb{R}^{l}$ that maps the obstacle set $\mathcal{O}_{it}$ to a finite vector describing the location, motion, and shape of each obstacle relative to the uncertain trajectory $c_{it}$. 
\subsection{Integrated Perception \& Motion Planning} 
We concatenate both the robot's state and the obstacle states at time $t$ to form the environmental state
\begin{align} \label{eqn_environment_state}
    \mathcal{Z}_t &= \begin{bmatrix} x_t \\  \mathcal{X}_{\mathcal{O}_{t}} \end{bmatrix} \in \mathbb{R}^{n+ln_0}, 
\end{align}
where $\mathcal{X}_{\mathcal{O}_{t}} = \begin{bmatrix}  \mathcal{X}_{1t} & \mathcal{X}_{2t} & \dots & \mathcal{X}_{n_{0}t} \end{bmatrix}^{\top}$ represents the concatenated states of all the $n_0$ obstacles at time $t$. Then the dynamics of the environmental state can be written as
\begin{align}\label{eqn_env_st_dynamics}
    \mathcal{Z}_{t+1} &= \underbrace{\begin{bmatrix} A & \mathbf{0}_{n \times ln_0} \\ \mathbf{0}_{ln_0 \times n} & I_{ln_0} \end{bmatrix}}_{A_{z}} \mathcal{Z}_{t} + \underbrace{\begin{bmatrix}B \\ \mathbf{0}_{ln_0 \times m} \end{bmatrix}}_{B_{z}} u_t + G_{z} \underbrace{\begin{bmatrix}w_{r,t} \\ w_{\mathcal{O},t} \end{bmatrix}}_{w_{t}},
\end{align}
where $G_{z} =$ diag$(G, I)$ and $w_{\mathcal{O}, t} \in \mathbb{R}^n$ is an obstacle process noise and can be derived from $c_{it}$. The distribution $P_{w_{t}}$ of $w_t$ is unknown and will be assumed to belong to an ambiguity set $\mathcal{P}^{w}$ of distributions satisfying
\begin{align} \label{eqn_ambig_set_w} 
    \mathcal{P}^{w} &= \{ P_{w_t} | \mathbb{E}[w_t] = 0, \mathbb{E}\left[ (w_t - \hat{w}_t)(w_t - \hat{w}_t)^{\top} \right]  = \Sigma_{w_t} \}.
\end{align}
At time $t$, the state of the robot can be extracted from the environmental state $\mathcal{Z}_{t}$ as
\begin{align}
    x_t &= \underbrace{\begin{bmatrix} \mathbf{1}_{n} & \mathbf{0}_{ln_0} \end{bmatrix}}_{C_{xr}} \mathcal{Z}_{t}.
\end{align}
In an autonomous robot, the environmental state $\mathcal{Z}_t$ must be estimated with a perception system from noisy on-board sensor measurements. We assume that a high-level perception system, such as Semantic \texttt{SLAM} described in \cite{semantic_slam}, processes high dimensional raw data $\Theta_{t} \in \mathbb{R}^N$ to recognize obstacles and produce noisy joint measurements of their state and the robot state. In particular, we define feature vectors $\mathcal{Y}_{xt} \in \mathbb{R}^{r}$ and $\mathcal{Y}_{it} \in \mathbb{R}^{q}$ for $i \in \mathcal{B}$ obtained through
\begin{align}
    \mathcal{Y}_{xt} &= \Upsilon_{x}(\Theta_t) \\
    \mathcal{Y}_{\mathcal{O}t} &= \Upsilon_{\mathcal{O}}(\Theta_t) = \begin{bmatrix} \mathcal{Y}_{1t} & \mathcal{Y}_{2t} & \dots & \mathcal{Y}_{n_{0}t} \end{bmatrix}^{\top},
\end{align}
where $\Upsilon_{x}: \mathbb{R}^{N} \rightarrow \mathbb{R}^{r}$ and $\Upsilon_{\mathcal{O}}: \mathbb{R}^{N} \rightarrow \mathbb{R}^{q n_0}$ are mappings defined by the \texttt{SLAM} algorithm to process the raw sensor data. We then represent these features as noisy measurements of the robot and obstacle states using an assumed linear (or linearized) output model
\begin{align} \label{eqn_output_sensor_model}
y_t &= \left[\begin{array}{c}y_{x,t} \\y_{\mathcal{O},t}\end{array}\right] = \mathcal{C}\left(\left[\begin{array}{c}\mathcal{Y}_{xt} \\\mathcal{Y}_{\mathcal{O}t}\end{array}\right]\right) = C \mathcal{Z}_t + H v_t, \\
C &= \begin{bmatrix}C_{r} \\ C_{\mathcal{O}}\end{bmatrix}, H = \begin{bmatrix}H_{r} \\ H_{\mathcal{O}}\end{bmatrix}, v_t = \begin{bmatrix}v_{r,t} \\ v_{\mathcal{O},t}\end{bmatrix} \sim P_{v_t}\in \mathcal{P}^v,
\end{align}
where $y_t \in \mathbb{R}^{p+s},$ and $y_{x,t} \in \mathbb{R}^p, y_{\mathcal{O},t} \in \mathbb{R}^s$ are the output vectors corresponding to the robot and the obstacles respectively. The matrices $C, C_{\mathcal{O}}, H_{x}, H_{\mathcal{O}}$ are of appropriate dimensions. The function $\mathcal{C}$ maps the feature vectors $\mathcal{Y}_{xt}, \mathcal{Y}_{\mathcal{O}t}$ to produce outputs $y_{xt}, y_{\mathcal{O}t}$ respectively as a linear function of the environmental state $\mathcal{Z}_t$ with additive measurement noises $v_t$ which is a zero-mean random variable. For simplicity, we assume that $w_t$ and $v_t$ are independent. The distribution $P_{v_{t}}$ of $v_t$ is assumed to belong to an ambiguity set, $\mathcal{P}^v$ satisfying
\begin{align}\label{eqn_ambig_set_v}
    \mathcal{P}^{v} &= \{ P_{v_t} | \mathbb{E}[v_t] = 0, \mathbb{E}\left[ (v_t - \hat{v}_t)(v_t - \hat{v}_t)^{\top} \right]  = \Sigma_{v_t} \}. 
\end{align}
 The robot is nominally subject to constraints on the state and input of the form, $\forall t = 0,\dots,T-1$,
\begin{align} \label{eqn_constraints}
x_t &\in \mathcal{X}^{\texttt{free}}_{t} = \mathcal{X} \backslash \bigcup_{i \in \mathcal{B}} \mathcal{O}_{it}, \\
u_t &\in \mathcal{U},
\end{align}
where the environment $\mathcal{X} \subset \mathbb{R}^{n}$, and $\mathcal{U} \subset \mathbb{R}^{m}$ are assumed to be
convex polytopes, The obstacles $\mathcal{O}_{it}, i \in \mathcal{B}$ are described by \eqref{eqn_obs_dynamics}, and the operator $\backslash$ denotes set subtraction. The set $\mathcal{B}$ represent a set of $n_0$ obstacles in the environment to be avoided.
\begin{equation}
    \begin{aligned}
        \mathcal{U} &= \{ u_t \, | \, \, A_{u} u_t \leq b_{u}\}, \\
        \mathcal{X} &= \{ \mathcal{Z}_{t} \, | \, \, A_{0} C_{xr} \mathcal{Z}_{t} \leq b_{0}\}, \\
        \mathcal{O}_{it} &= \{ \mathcal{Z}_{t} \, | \, \, A_{i} C_{xr} \mathcal{Z}_{t} \leq b_{it}\}, \quad i \in \mathcal{B}\\
    \end{aligned}
\end{equation}
where $b_u \in \mathbb{R}^{n_u}, b_0 \in \mathbb{R}^{n_E}, b_{it} \in \mathbb{R}^{n_i},$ and $A_u, A_0,$ and $A_i$ are matrices of appropriate dimension. The nonconvex obstacle avoidance constraints for obstacle $i \in \mathcal{B}$ can be expressed as the disjunction
\begin{align}
    \neg (A_{i} C_{xr} \mathcal{Z}_{t} \leq b_{it}) \Leftrightarrow \bigvee^{n_i}_{j=1} (a^{\top}_{ij} C_{xr} \mathcal{Z}_{t} \geq a^{\top}_{ij} c_{it}),
\end{align}
where $\lor$ denotes disjunction.
\subsection{A Distributionaly Robust Motion Planning Problem}
We seek a dynamic output feedback control policy $\pi = [\pi_0,\dots, \pi_{T-1}]$ with $u_t = \pi_t(y_{0:t}, u_{0:t-1})$, where $y_{0:t}$ and $u_{0:t-1}$ are the output and input histories available to make control decisions at time $t$, that produces a feasible and minimum cost trajectory from an initial state $x_0$ to a goal set $\mathcal{X}_{goal} \subset \mathbb{R}^{n}$. In particular, we seek to (approximately) solve the distributionally robust constrained stochastic optimal control problem
\begin{equation} \label{eqn_dr_motion_planning}
    \begin{aligned}
        &\underset{\pi}{\text{minimize}} & & \sum^{T-1}_{t = 0} \ell_t(\mathbb{E}\left[\mathcal{Z}_t\right], \mathcal{X}_{goal}, u_t) + \ell_T(\mathbb{E}\left[\mathcal{Z}_T\right], \mathcal{X}_{goal}) \\
        &\text{subject to } & &  \mathcal{Z}_{t + 1} = A_{z} \mathcal{Z}_{t} + B_{z} u_t + G_{z} w_t, \\
        & & & y_{t} = C \mathcal{Z}_t + H v_t \\
        & & & \mathcal{Z}_0 \sim P_{\mathcal{Z}_0}\in \mathcal{P}^\mathcal{Z}, \\
        & & & w_t \sim P_{w}\in \mathcal{P}^w, \\
        & & & u_t \in \mathcal{U} = \{ u_t \, | \, \, A_{u} u_t \leq b_{u}\},  \\
        & & & \mathcal{X}^{\texttt{free}}_{t} = \mathcal{X} \backslash \bigcup_{i \in \mathcal{B}} \mathcal{O}_{it}, \\
        & & & \underset{P_{\mathcal{Z}_t}\in \mathcal{P}^{\mathcal{Z}}}{\inf} P_{\mathcal{Z}_t} (C_{xr} \mathcal{Z}_{t} \in \mathcal{X}^{\texttt{free}}_{t}) \geq 1 - \alpha, 
    \end{aligned}
\end{equation}
where $\mathcal{P}^\mathcal{Z}$ is an ambiguity set of marginal state distributions and $\alpha \in (0,0.5]$ is a user-prescribed risk parameter. The stage cost functions $\ell_t(\cdot)$ quantify the robot’s distance to the goal set and actuator effort, and are assumed to be expressed in terms of the environmental state mean $\mathbb{E}[\mathcal{Z}_t]$, so that all the stochasticity appears in the constraints. Two key features distinguish our problem formulation. First, the state constraints are expressed as \emph{distributionally robust chance constraints}. This means that the nominal constraints $x_t \in \mathcal{X}^{\texttt{free}}_{t}$ are enforced with probability $\alpha$ under the worst-case distribution in the ambiguity set. Second, since information about the environmental state is obtained only from noisy measurements, we optimize over dynamic output feedback policies. Our proposed solution framework, detailed in the next section, combines a dynamic state estimator with a full-state kinodynamic motion planning under uncertainty algorithm. This combination and explicit incorporation of state estimation uncertainty into the motion planning and control takes a step toward tighter integration of perception, planning, and control, which are nearly always separated in state-of-the-art robotic systems.

\section{Output Feedback Distributionally Robust $\rrtstar$ (\texttt{OFDR}-$\rrtstar$)} \label{sec_dr_rrt}
We propose to use a distributionally robust, kinodynamic variant of the $\rrtstar$ motion planning algorithm with dynamic output feedback policies. $\rrtstar$ adds a rewiring operation to \texttt{RRT} to obtain asymptotic optimality. Our proposed algorithm grows trees of state and state estimate distributions, rather than merely trees of states, and incorporates distributionally robust probabilistic constraints to build risk-constrained state trajectories and feedback policies. 

\subsection{LQG Control Based Steering Law} \label{subsec_steer}
Sampling based motion planning algorithms require a steering law to steer the robot from a noide in the tree to a feasible sampled point in the free space. Since the environment state is not directly observed and must be estimated from noisy output measurements \eqref{eqn_output_sensor_model}, our proposed steering law $\pi = [\pi_0,\dots, \pi_{T-1}]$ with $u_t = \pi_t(y_{0:t}, u_{0:t-1})$ comprises a combination of dynamic state estimator and state feedback control law. Here we utilize a Kalman filter (which has been used in seminal \texttt{SLAM} algorithms for joint estimation of robot and environmental states \cite{dissanayake2001solution}) for state estimation together with a finite horizon optimal linear quadratic state feedback controller. It is also possible within our framework to more sophisticated estimation and control components (e.g., extended/unscented Kalman filters or particle filters and stochastic model predictive controllers), which will be explored in future work. 

The output feedback control policy has the form
\begin{align} \label{eqn_affine_control_law_with_KF}
    u_t = K_t \tilde{\mathcal{Z}}_t + k_t.
\end{align}
where $\tilde{\mathcal{Z}}_t$ is the Kalman filter estimate of the environmental state, and $K_t$ and $k_t$ are linear and constant feedback gains to be derived with dynamic programming for a finite-horizon LQR problem.

\textbf{Kalman Filter:} The Kalman filter equations with gain $\mathcal{L}_{t}$ are
\begin{align}
    \mathcal{L}_{t+1} &= {\Sigma}_{\tilde{\mathcal{Z}}_{t}} C^{\top} \left( C {\Sigma}_{\tilde{\mathcal{Z}}_{t}} C^{\top} + H \Sigma_{v_t} H^{\top} \right)^{-1}, \\
    \tilde{\mathcal{Z}}_{t+1} &= (I - \mathcal{L}_{t+1} C) (A_{z} \tilde{\mathcal{Z}}_t + B_{z} u_t) + \mathcal{L}_{t+1} y_{t+1}, \\
     \Sigma_{\tilde{\mathcal{Z}}_{t+1}} &= (I - \mathcal{L}_{t+1} C) (A_{z} \Sigma_{\tilde{\mathcal{Z}}_{t}} A^{\top}_{z} + G_{z}  \Sigma_{w_t} G^{\top}_{z}).
\end{align}
Together with the control law \eqref{eqn_affine_control_law_with_KF} we can write the combined dynamics for the true unknown state $\mathcal{Z}_t$ and the state estimate $\tilde{\mathcal{Z}}_t$ as
\begin{align}
    \underbrace{\begin{bmatrix} \mathcal{Z}_{t+1} \\ \tilde{\mathcal{Z}}_{t+1} \end{bmatrix}}_{Z_{t+1}} &= \underbrace{\begin{bmatrix} A_{z} & B_{z} K_t \\ \mathcal{L}_{t+1} C A_{z} & (I - \mathcal{L}_{t+1} C) A_{z} + B_{z} K_t\end{bmatrix}}_{\bar{A}_{t}} \underbrace{\begin{bmatrix} \mathcal{Z}_{t} \\ \tilde{\mathcal{Z}}_{t} \end{bmatrix}}_{Z_{t}} \\
    &+ \underbrace{\begin{bmatrix} B_{z} \\ B_{z} \end{bmatrix}}_{\bar{B}_{t}} \begin{bmatrix} k_{t} \end{bmatrix} + \underbrace{\begin{bmatrix} G_{z} & 0 \\ \mathcal{L}_{t+1} C G_{z} & \mathcal{L}_{t+1} H \end{bmatrix}}_{\bar{G}_{t}} \underbrace{\begin{bmatrix} w_{t} \\ v_{t+1} \end{bmatrix}}_{W_{t}}, 
\end{align}
Let us define $\bar{Z}_t = Z_t - \hat{Z}_t$, $\bar{W}_t = W_t - \hat{W}_t$. Then,
\begin{align}
 \mathbb{E} \left[\bar{Z}_t \bar{Z}^{\top}_t \right] &= \Pi_{t} = \tilde{A}_{t-1} \Psi_{t-1} \tilde{A}^{\top}_{t-1} \\
 \mathbb{E} \left[\bar{W}_t \bar{W}^{\top}_t \right] &= \Sigma_{W_t} = \begin{bmatrix} \Sigma_{w_t} & 0 \\ 0 & \Sigma_{v_{t+1}} \end{bmatrix}
\end{align}
where $\tilde{A}_{t} = \begin{bmatrix} \bar{A}_t & \bar{G}_{t} W_t \end{bmatrix}, \Psi_{t} = \begin{bmatrix}\bar{Z}_t \bar{Z}^{\top}_t & \bar{Z}_t \\ \bar{Z}^{\top}_t & I \end{bmatrix}$. With the initial estimates $\hat{Z}_{0} = \begin{bmatrix} \hat{Z}_0 \\ \hat{Z}_0 \end{bmatrix},\Pi_{0} = \begin{bmatrix} \Sigma_{Z_{0}} & 0 \\ 0 & 0 \end{bmatrix}$, $\Sigma_{W_{0}} = \begin{bmatrix} \Sigma_{w_{0}} & 0 \\ 0 & \Sigma_{v_{0}} \end{bmatrix}$ given (or estimated from historical data), the combined state and state estimate mean and covariance evolve as
\begin{align}
    \Hat{Z}_{t+1} &= \bar{A}_t \Hat{Z}_{t} + \bar{B}_t k_t, \label{eqn_Z_mean_extract}\\
    \Pi_{t+1} &= \bar{A}_t \Pi_{t} \bar{A}^{\top}_t + \bar{G}_{t} \Sigma_{W_{t}} \bar{G}^{\top}_{t}. \label{eqn_Z_cov_extract}
\end{align}
For analysis purposes, the unknown environmental state mean and covariance can then be extracted as
\begin{align}
    \mathcal{Z}_{t+1} &= \begin{bmatrix} \mathbf{1}^{\top} & \mathbf{0}^{\top}  \end{bmatrix} \Hat{Z}_{t+1}, \\
    \Sigma_{\mathcal{Z}_{t+1}} &= \begin{bmatrix} \mathbf{I}_n \\ \mathbf{0}_n \end{bmatrix}^{\top} \Pi_{t+1} \begin{bmatrix} \mathbf{I}_n \\ \mathbf{0}_n \end{bmatrix}. 
\end{align}
\textbf{Optimal Finite-Horizon LQR Control:} Define the error $e_t = C_{xr}\tilde{\mathcal{Z}}_{t} - x_{s}$, where $x_s$ represents a sample of the free space to be steered to. Then the optimal control gains in \eqref{eqn_affine_control_law_with_KF} can be obtained by minimizing the cost function 
\begin{align} \label{eqn_sim_cost_fn}
    J &= \mathbb{E} \left[ \sum^{T-1}_{t=0} \left(e^{\top}_{t} Q e_{t} + u^{\top}_{t} R u_t \right) + e^{\top}_{T} Q e_{T} \right], 
\end{align}
via dynamic programming with the following backward in time recursion from $t = T,\dots,1$
\begin{align} \label{eqn_lqr_recursion}
    K_t &= - \beta^{-1} B^{\top}_{z} P_{t+1} A_{z}, \\ 
    k_t &= - \beta^{-1} B^{\top}_{z} q_{t+1}, \\
    P_t &= Q + A^{\top}_{z} P_{t+1} \left[ I - B_{z} \beta^{-1} B^{\top}_{z} P_{t+1} \right] A_{z}, \\
    q_t &= (A_{z} + B_{z}K_t)^{\top} q_{t+1} +  (K_t \beta + A^{\top}_{z} P_{t+1} B_{z}) k_{t} - Q x_s,
\end{align}
with initial values $P_T = Q$, $q_T = - Q x_s$ and further $\beta = (R+B^{\top}_{z} P_{t+1}B_{z})$ and $Q, R$ are the state and control cost matrices, respectively. 

\subsection{Moment-Based Ambiguity Set To Model Uncertainty}
Unlike most stochastic motion planning algorithms that often assume a functional form (often Gaussian) for probability distributions to model uncertainties, we will focus here on uncertainty modeling using moment-based ambiguity sets. Based on the ambiguity sets for the primitive random variables (namely, the process noise $w$ in \eqref{eqn_ambig_set_w}, the measurement noise $v$ in \eqref{eqn_ambig_set_v}, and the analogous one for the initial state and state estimate $Z_0$), and based on the estimator and control law, the combined environmental state and state estimate and covariance propagate according to \eqref{eqn_Z_mean_extract} and \eqref{eqn_Z_cov_extract}. Since the primitive distributions are not assumed to be Gaussian, then neither are the marginal state and state estimate distributions distributions $P_{Z_{t}}$. The ambiguity set defining the combined environmental state and state estimate is 
\begin{align} 
    \mathcal{P}^{Z} &= \left\{ P_{Z_t} \, | \, \, \mathbb{E}[Z_t] = \hat{Z}_{t}, \mathbb{E} [(Z_t - \hat{Z}_t)(Z_t - \hat{Z}_t)^{\top}] = \Sigma_{Z_t} \right\},
\end{align}
and the ambiguity set for the true environmental state is
\begin{align} \label{eqn_ambiguity_set_env_state}
    \mathcal{P}^{\mathcal{Z}} &= \left\{ P_{\mathcal{Z}_t} \, | \, \, \mathbb{E}[\mathcal{Z}_t] = \hat{\mathcal{Z}}_{t}, \mathbb{E} [(\mathcal{Z}_t - \hat{\mathcal{Z}}_t)(\mathcal{Z}_t - \hat{\mathcal{Z}}_t)^{\top}] = \Sigma_{\mathcal{Z}_t} \right\}.
\end{align}

\subsection{Distributionally Robust Collision Check}
The control law returned by the steering function should also satisfy the state constraints which are expressed as distributionally robust chance constraints. In particular, the nominal state constraints, $C_{xr} \mathcal{Z}_t \in \mathcal{X}^{\texttt{free}}_{t}$, are required to be satisfied with probability $1 - \alpha$, under the worst case probability distribution in the ambiguity set. Let the moment-based ambiguity set for obstacle motion be defined using $\mathbf{E}[c_{it}] = \hat{c}_{it}$ and $\mathbf{E}[(c_{it} - \hat{c}_{it})(c_{it} - \hat{c}_{it})^{\top}] = \Sigma^{c}_{it}$. Now, under the moment-based ambiguity set defined by \eqref{eqn_ambiguity_set_env_state}, a constraint on the worst-case probability of violating the $j^{th}$ constraint of obstacle $i \in \mathcal{B}$
\begin{equation} \label{eqn_dr_constraint}
    \underset{P_{\mathcal{Z}_t} \in \mathcal{P}^{\mathcal{Z}}}{\sup} P_{\mathcal{Z}_t}(a^{\top}_{ij} C_{xr} \mathcal{Z}_t \geq a^{\top}_{ij} c_{it}) \leq \alpha_i
\end{equation}
is equivalent to the linear constraint on the state mean $\hat{\mathcal{Z}}_t$
\begin{equation} \label{eqn_dr_constraint_tightening}
    a^{\top}_{ij} C_{xr} \hat{\mathcal{Z}}_t \geq a^{\top}_{ij} \hat{c}_{it} + \sqrt{\frac{1 - \alpha_i}{\alpha_i}} {\left \| (D_{\hat{x}_{t}} + \Sigma^{c}_{it})^{\frac{1}{2}} a_{ij} \right \Vert}_{2},
\end{equation}
where $D_{\hat{x}_{t}} = C_{xr} \Sigma_{\mathcal{Z}_{t}} C^{\top}_{xr}$ and $\alpha_{i}$ is the user prescribed risk parameter for obstacle $i \in \mathcal{B}$. Obstacle risks are allocated such that their sum does not exceed the constraint risk $\alpha$. The scaling constant $\sqrt{\frac{1 - \alpha_i}{\alpha_i}}$ in the deterministic tightening of the nominal constraint is larger than the Gaussian one, leading to a stronger tightening that reflects the weaker assumptions about the uncertainty distributions. 

\subsection{Sample-Based Motion Planning Algorithm} \label{subsec_drrrt_algorithm}
The \texttt{OFDR}-$\rrtstar$ tree expansion procedure, similar to the \texttt{CC}-$\rrtstar$ algorithm developed in \cite{luders_rrtstar}, is presented in Algorithm \ref{alg_tree_expansion}. The \texttt{OFDR}-$\rrtstar$ tree is denoted by $\mathcal{T}$, consisting of $|\mathcal{T}|$ nodes. Each node $N$ of the tree $\mathcal{T}$ consists of a sequence of state distributions, characterized by a distribution mean $\hat{x}$ and covariance
$D$. A sequence of means and covariances is denoted by $\bar{\sigma}$ and $\bar{\Pi}$, respectively. The final mean and covariance of a node's sequence are denoted by $x[N]$ and $D[N]$, respectively. For the state distribution sequence $(\bar{\sigma},\bar{\Pi})$, the notation $\Delta J(\bar{\sigma},\bar{\Pi})$ denotes the cost of that sequence. If $(\bar{\sigma},\bar{\Pi})$ denotes the trajectory of node $N$ with parent $N_{parent}$, then we denote by $J[N]$, the entire path cost from the starting state to the terminal state of node $N$, constructed recursively as
\begin{align} \label{eqn_node_cost}
    J[N] = J[N_{parent}] + \Delta J(\bar{\sigma},\bar{\Pi}).
\end{align}
In the first step, a random sample $x_{rand}$ is taken from the feasible state set, $\mathcal{X}^{\texttt{free}}_{t}$. Then the tree node, $N_{nearest}$ that is nearest to the sample is selected (line 3 of Algorithm \ref{alg_tree_expansion}) according to an optimal cost-to-go function without the obstacle constraints, in order to efficiently explore the reachable set of the dynamics and increase the likelihood of generating collision-free trajectories (similar to what is done in \cite{frazzoli_real_mp}). Attempts are then made to steer the robot from the nearest tree node to the random sample using the steering law explained in subsection \ref{subsec_steer} (line 4). The control policy obtained is then used to propagate the state mean and covariance, and the entire trajectory $(\bar{\sigma},\bar{\Pi})$ is returned by the steer function. Each state distribution in the trajectory is then checked for distributionally robust probabilistic constraint satisfaction given by \eqref{eqn_dr_constraint_tightening} and further the line connecting subsequent state distributions in the trajectory are also checked for collision with the obstacle sets $\mathcal{O}_{it}, i \in \mathcal{B}$. An outline of the \texttt{DR-Feasible} subroutine is shown in Algorithm \ref{alg_dr_feasible_module}. If the entire trajectory $(\bar{\sigma},\bar{\Pi})$ is probabilistically feasible, a new node $N_{min}$ with that distribution sequence $(\bar{\sigma},\bar{\Pi})$ is created (line 7) but not yet added to $\mathcal{T}$. Instead, nearby nodes are identified for possible connections via the NearNodes function (line 8), which returns a subset of nodes $\mathcal{N}_{near} \subseteq \mathcal{T}$, if they are within a search radius ensuring probabilistic asymptotic optimality guarantees specified in \cite{luders_rrtstar}
\begin{align}
    r = \min\left\{ \gamma \left( \frac{\log(|\mathcal{N}_{t}|)}{\mathcal{N}_{t}}  \right)^{1/d}, \mu_{max} \right\}, 
\end{align}
where $\mathcal{N}_{t}$ refers to the number of nodes in the tree at time $t$, $\mu_{max} > 0$ is the maximum radius specified by the user, $\gamma$ refers to the planning constant based on the $d$ dimensional environment. Then we seek to identify the lowest-cost, probabilistically feasible connection from the $\mathcal{N}_{near}$ nodes to $x_{rand}$ (lines 10-14). For each possible connection, a distribution sequence is simulated via the steering law (line 11). If the resulting sequence is probabilistically feasible, and the cost of that node represented as $c_{rand} = J[N_{near}] + \Delta J(\bar{\sigma},\bar{\Pi}$, is lower than the cost of $N_{min}$ denote by $J[N_{min}]$, then a new node with this sequence replaces $N_{min}$ (line 14). The lowest-cost node is ultimately added to $\mathcal{T}$ (line 15). 

Finally, edges are rewired based on attempted connections from the new node $N_{min}$ to nearby nodes $\mathcal{N}_{near}$ (lines 17-22), ancestors excluded (line 17). A distribution sequence is simulated via the steering law from $N_{min}$ to the terminal state of each nearby node $N_{near} \in \mathcal{N}_{near}$ (line 18). If the resulting sequence is probabilistically feasible, and the cost of that node $c_{near}$ is lower than the cost of $N_{near}$ given by $J[N_{near}]$ (line 19), then a new node with this distribution sequence replaces $N_{near}$ (lines 21-22). The tree expansion procedure is then repeated until a node from the goal set is added to the tree. At that point, a distributionally robust feasible trajectory is obtained from the tree root to $\mathcal{X}_{goal}$.
\begin{algorithm}
\caption{\texttt{\texttt{OFDR}-$\rrtstar$}- Tree Expansion Procedure} \label{alg_tree_expansion}
\begin{algorithmic}[1]
\State $\text{Inputs: Current Tree } \mathcal{T}, \text{ time step } t$
\State $x_{rand} \gets \text{Sample} (\mathcal{X}_t)$
\State $N_{nearest} \gets$ NearestNode$(x_{rand}, \mathcal{T})$
\State $(\bar{\sigma}, \bar{\Pi}) \gets$ Steer$(\hat{x}[N_{nearest}], D[N_{nearest}], x_{rand})$
\State \text{\color{gray} // \texttt{Check if sequence} $(\bar{\sigma}, \bar{\Pi})$ \texttt{is DR-Feasible}}
\If{\texttt{DR-Feasible}$(\bar{\sigma}, \bar{\Pi})$} 
\State Create node $N_{min}\{\bar{\sigma}, \bar{\Pi}\}$
\State $\mathcal{N}_{near} \gets$ NearNodes$(\mathcal{T}, x_{rand}, \left | \mathcal{T} \right \vert)$
\State \text{\color{gray} // \texttt{Connect along a minimum-cost path}}
\ForEach {$N_{near} \in \mathcal{N}_{near} \backslash N_{nearest}$}
\State $(\bar{\sigma}, \bar{\Pi}) \gets$ Steer$(\hat{x}[N_{near}], D[N_{near}], x_{rand})$
\State $c_{rand} \gets$ J$[x_{near}] + \Delta J(\bar{\sigma}, \bar{\Pi})$
\If{\texttt{DR-Feasible}$(\bar{\sigma}, \bar{\Pi})$ \& $c_{rand} < J[N_{min}]$}
\State Replace $N_{min}$ with $N_{min}\{\bar{\sigma}, \bar{\Pi}\}$
\EndIf
\EndFor
\State Add $N_{min}$ to $\mathcal{T}$
\State \text{\color{gray} // \texttt{Re-Wire the Tree}}
\ForEach {$N_{near} \in \mathcal{N}_{near} \backslash$ Ancestors$(N_{min})$}
\State $(\bar{\sigma}, \bar{\Pi}) \gets$ Steer$(\hat{x}[N_{min}], D[N_{min}], \hat{x}[N_{near}])$
\State $c_{near} \gets$ J$[N_{min}] + \Delta J(\bar{\sigma}, \bar{\Pi})$
\If{\texttt{DR-Feasible}$(\bar{\sigma}, \bar{\Pi})$ \& $c_{near} <  J[N_{near}]$}
\State Delete $N_{near}$ from $\mathcal{T}$
\State Add new node $N_{new}\{\bar{\sigma}, \bar{\Pi}\}$ to $\mathcal{T}$
\EndIf
\EndFor
\EndIf
\end{algorithmic}
\end{algorithm}

\begin{algorithm}
\caption{\texttt{DR-Feasible} Subroutine}\label{alg_dr_feasible_module}
\begin{algorithmic}[2]
\State $\text{Input: } \mathbf{T-} \text{time step distribution sequence} (\bar{\sigma}, \bar{\Pi})$
\For{$t = 1$ to $\mathbf{T}$}
\State $(\hat{x}_t, D_{\hat{x}_{t}}) \gets$ $t\ts{th}$ element in $(\bar{\sigma}, \bar{\Pi})$ sequence
\State $\mathbb{L} \gets$ Line connecting position block of $\hat{x}_{t-1}$ to
$\hat{x}_{t}$
\ForEach{$i \in \mathcal{B}$}
\If{$(\hat{x}_t, D_{\hat{x}_{t}})$ dissatisfies \eqref{eqn_dr_constraint_tightening} or $\mathbb{L} \in \mathcal{O}_{it}$}
\State Return \texttt{false} 
\EndIf
\EndFor
\EndFor
\State Return \texttt{true}
\end{algorithmic}
\end{algorithm}

\section{Simulation Results} \label{sec_sim_results}
We demonstrate our proposed framework using a double integrator robot moving in a bounded and cluttered environment. While the proposed framework can handle dynamic and uncertain obstacles, for simplicity of illustration we assume the obstacles are static and deterministic $(w_{\mathcal{O}_{t}}=0, \forall t)$, so that all uncertainty in this example comes from the unknown initial state, robot process disturbance, and measurement noise. The robot dynamics matrices are 
\begin{align*} \label{eqn_robot_matrices}
    A &= \begin{bmatrix} 1 & 0 & dt & 0 \\ 0 & 1 & 0 & dt \\ 0 & 0 & 1 & 0 \\ 0 & 0 & 0 & 1 \end{bmatrix}, B = \begin{bmatrix} \frac{{dt}^2}{2} & 0 \\ 0 & \frac{{dt}^2}{2} \\ dt & 0 \\ 0 & dt \end{bmatrix}, G = B
\end{align*}
where $dt = 0.1s$ and the states are the two dimensional position and velocity with two dimensional force inputs. The environmental state dynamics matrices $A_z, B_z, G_z$ are formed accordingly using \eqref{eqn_env_st_dynamics} with the above robot dynamics matrices. We assume the robot to start from the origin with zero initial velocity, and that the initial state and noise covariance matrices are 
\begin{align*}
    \Sigma_{x_{0}} &= 0.1 \begin{bmatrix} 1 & 0 & 0 & 0 \\ 0 & 1 & 0 & 0 \\ 0 & 0 & 1 & 0 \\ 0 & 0 & 0 & 1 \end{bmatrix}, \Sigma_{w} = 0.1 \begin{bmatrix} 0 & 0 & 0 & 0 \\ 0 & 0 & 0 & 0 \\ 0 & 0 & 2 & 1 \\ 0 & 0 & 1 & 2 \end{bmatrix}, \Sigma_{v} = 10^{-3} I.
\end{align*}
The robot is treated as a point mass without loss of generality, since a known geometry can be easily handled by a fixed tightening of the state constraints. The 2D position of the robot is sampled uniformly within the bounds of the feasible 2D environment whose boundaries are not treated probabilistically. The search radius used in the Algorithm \ref{alg_tree_expansion} uses a maximum radius of $\mu_{max} = 1$ and the environment based planning constant $\gamma = 20$. The output filter dynamics matrices are
\begin{align}
    C_{r} &= \begin{bmatrix} 1 & 0 & 0 & 0 \\ 0 & 1 & 0 & 0 \end{bmatrix}, C_{\mathcal{O}} = I_{s \times ln_0}, H_r = I, H_{\mathcal{O}} = \mathbf{0}.
\end{align}
We define the error $e_{t} = C_{xr}\mathcal{Z}_{t} - x_{s}$ for $i = 0,\dots,T$, with $T=5$. A dynamic output feedback policy $u_t$ of the form given by \eqref{eqn_affine_control_law_with_KF} that minimizes the cost function 
\begin{align} \label{eqn_cost_function}
    J &= \mathbb{E} \left[ \sum^{T-1}_{t=0} \left(e^{\top}_{t} Q e_{t} + u^{\top}_{t} R u_t \right) + e^{\top}_{T} Q e_{T} \right], 
\end{align}
is computed using dynamic programming to steer the robot from a tree node state $x_t$ to a random feasible sample $x_s$. The matrices $Q = \begin{bmatrix} 40I & 0 \\ 0 & 2I \end{bmatrix}, R = 0.02 I$ are used to penalize the state and control deviations respectively. The distributionally robust state constraints are enforced with probabilistic satisfaction parameter $\alpha = 0.05$. Three different simulations using the above double integrator system are performed namely: 1) deterministic collision check where uncertainties are not accounted for, 2) chance constrained collision check where the system noises are assumed to be Gaussian distributed and, 3) distributionally robust collision check assuming the noises belong to their respective ambiguity sets. For all the simulations, the chance constrained $\rrtstar$ algorithm with the corresponding collision check procedure is run for 1200 iterations, with 1-$\sigma$ position uncertainty ellipses from the covariance matrix being drawn at the end of each trajectory. 
\subsection{Results and Discussion}
\begin{figure}
    \centering
    \includegraphics[scale=0.20]{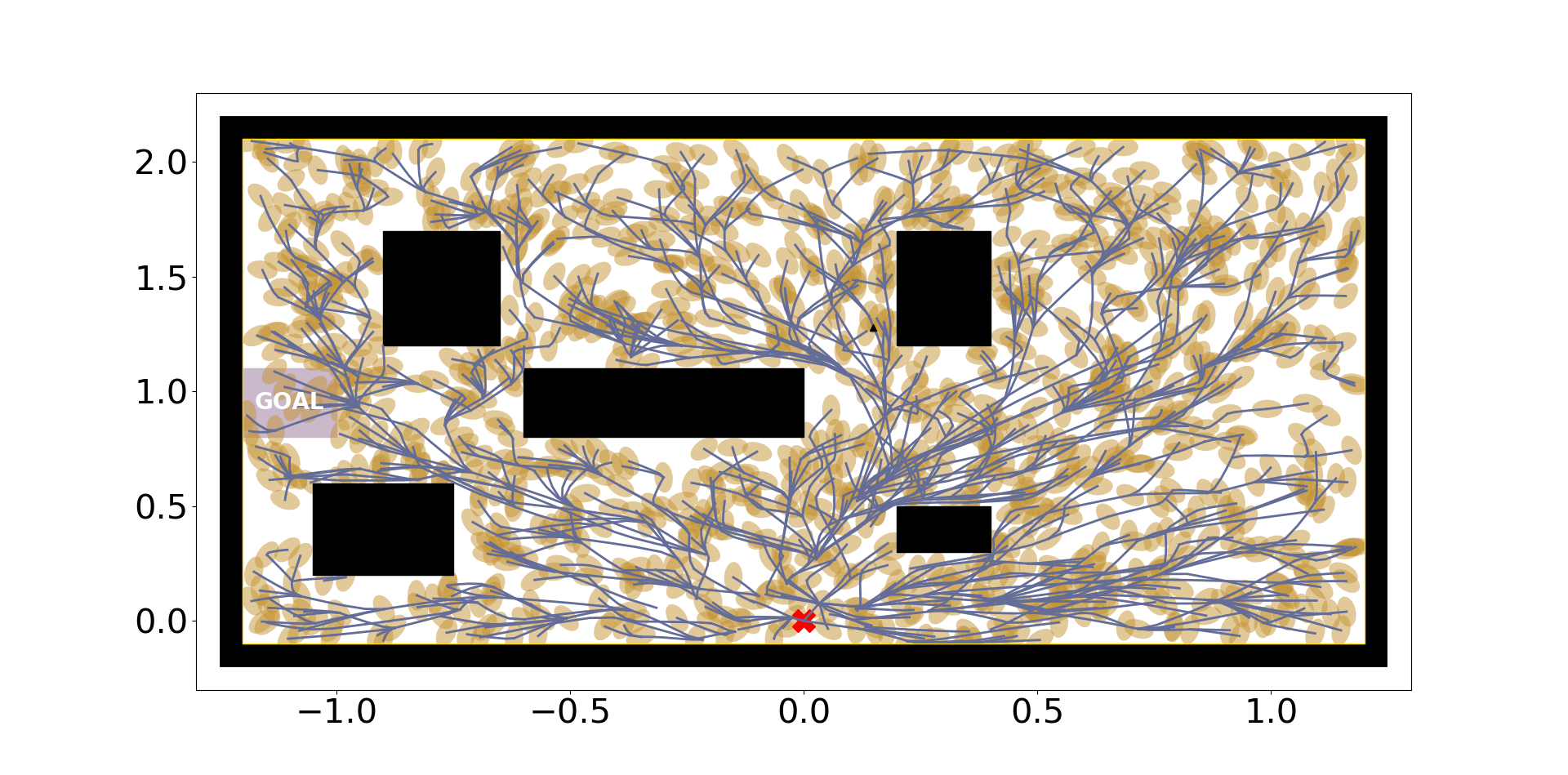}
    \caption{\textbf{No Risk Constraints.} An $\rrtstar$ tree of 1200 nodes showing nominal states and 1-$\sigma$ position uncertainty ellipses, generated without any constraint tightening. Since uncertainty is not explicitly incorporated into collision checking, it produces in highly risky trajectories.}
    \label{fig_det_rrt_star_env}
\end{figure}
\begin{figure}
    \centering
    \includegraphics[scale=0.20]{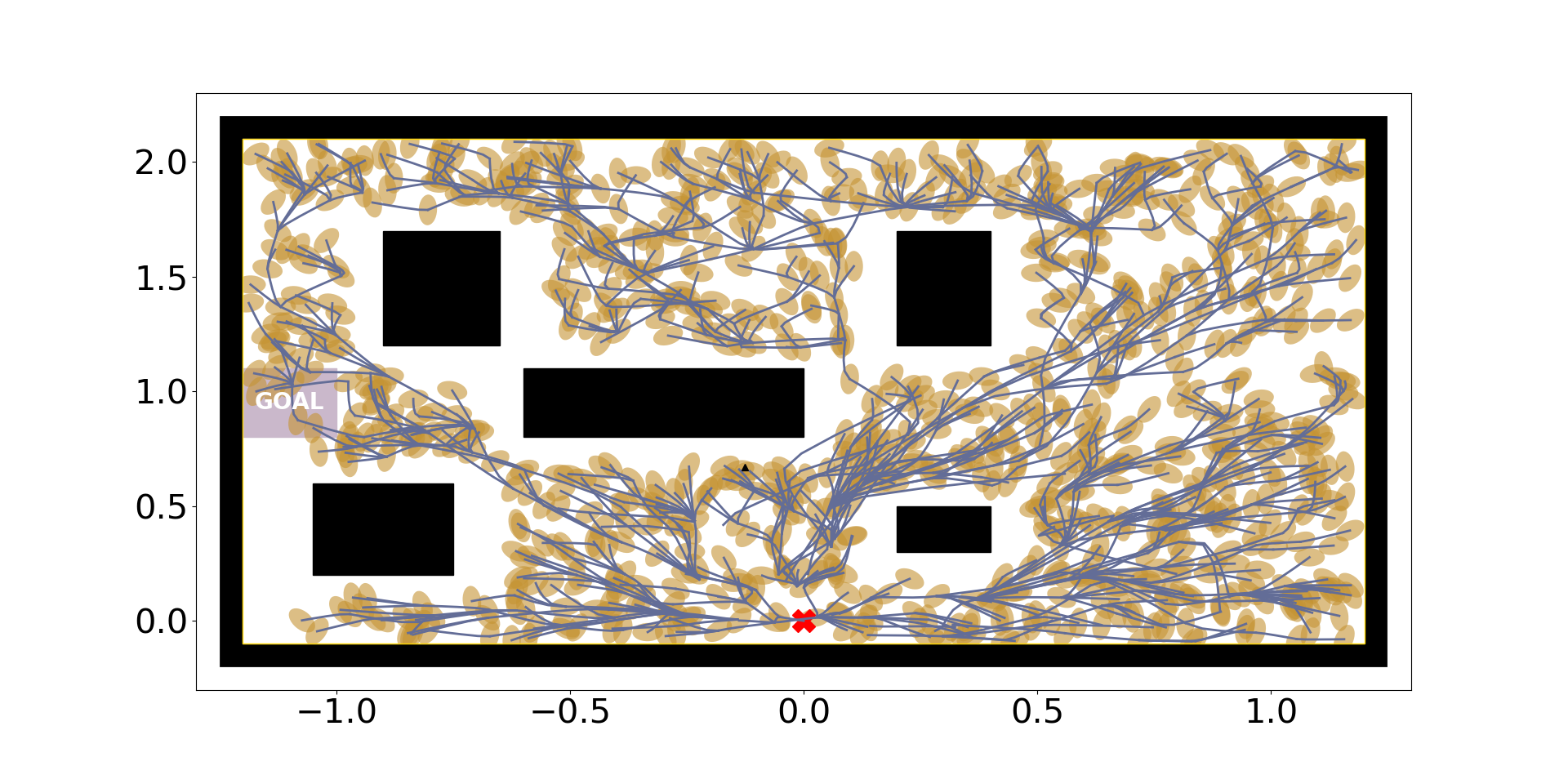}
    \caption{\textbf{Gaussian Risk Constraints.} An $\rrtstar$ tree of 1200 nodes showing nominal states and 1-$\sigma$ position uncertainty ellipses, generated with Gaussian chance constraints with $\alpha = 0.05$. Although the trajectories are less risky than when uncertainty is ignored, this approach can still significantly underestimate risks of constraint violation since the actual perception uncertainties in robotic systems often do not match well with a Gaussian assumption.}
    \label{fig_cc_rrt_star_env}
\end{figure}
\begin{figure}
    \centering
    \includegraphics[scale=0.20]{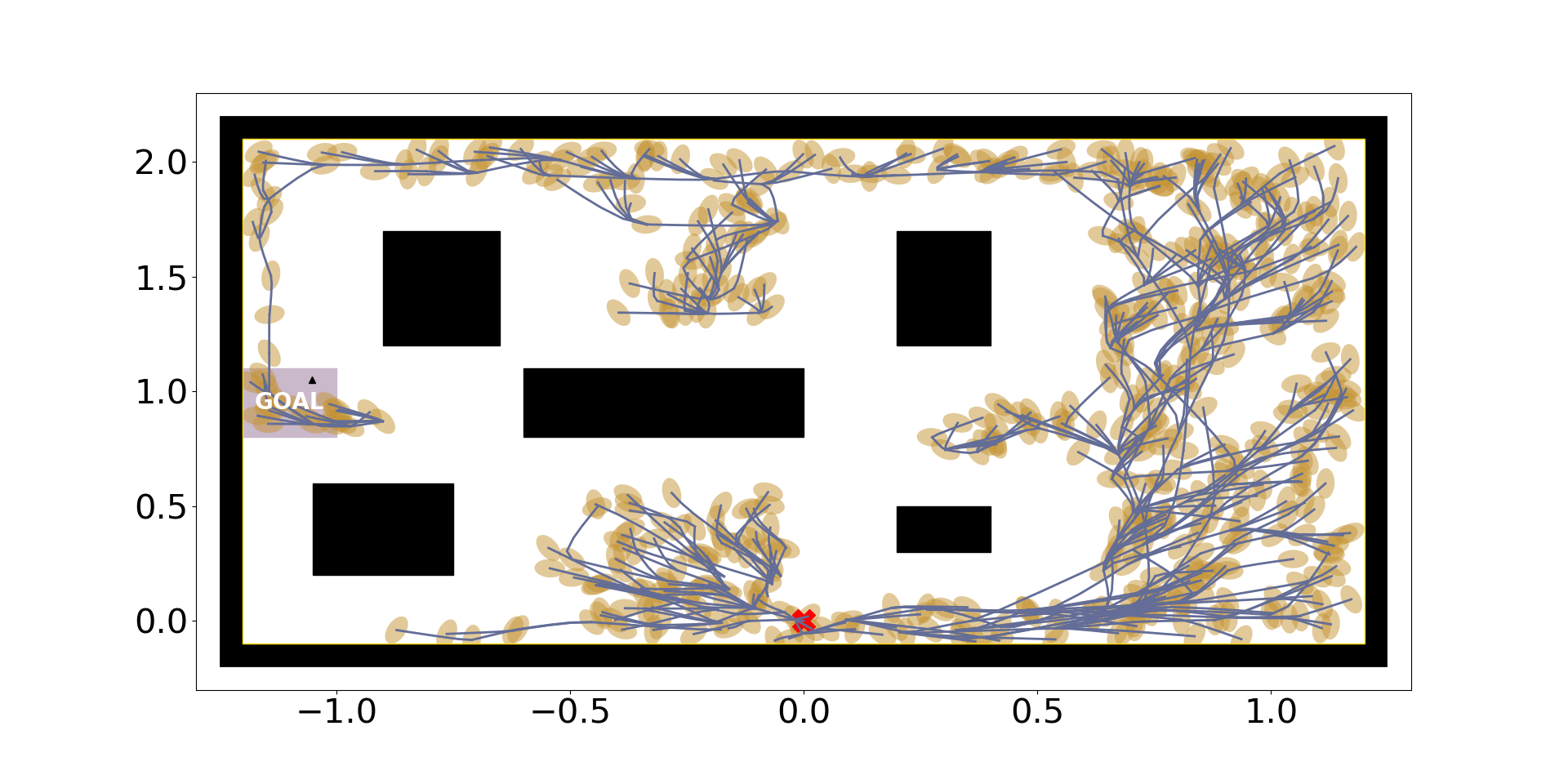}
    \caption{\textbf{\texttt{DR} Risk Constraints.} An $\rrtstar$ tree of 1200 nodes showing nominal states and 1-$\sigma$ position uncertainty ellipses, generated with \texttt{OFDR}-$\rrtstar$ using distributionally robust chance constraints with $\alpha = 0.05$. This approach produces more risk-averse trajectories with a more sophsticated and coherent risk quantification. }
    \label{fig_dr_rrt_star_env}
\end{figure}
The tree generated by an $\rrtstar$ algorithm using a deterministic collision check where uncertainties are not accounted for is shown in Figure \ref{fig_det_rrt_star_env}. It can be seen that highly risky trajectories around the obstacles are generated, since the uncertainty in the state due to the initial localization and system dynamics uncertainties are not explicitly incorporated. Assuming Gaussian noises and using risk parameter $\alpha = 0.05$, the chance constrained variant of $\rrtstar$ generates more conservative trajectories as shown in Figure \ref{fig_cc_rrt_star_env}. However, the actual perception uncertainties in robotic systems often do not match well with a Gaussian assumption, so this approach can still significantly underestimate risks of constraint violation. 

The \texttt{OFDR}-$\rrtstar$ tree as shown in Figure \ref{fig_dr_rrt_star_env} generates more conservative trajectories around the obstacles than the Gaussian chance constrained counterpart, by explicitly incorporating the uncertainty in the state due to the initial localization, system dynamics, and measurement uncertainties in the form of ambiguity sets. It produces trajectories that satisfy the chance constraints under the worst-case distribution in the ambiguity sets. Clearly, the feasible set is smaller with the distributionally robust constraints, and certain nominally feasible paths from the initial state to the goal are deemed too risky in the presence of the uncertainties, e.g., the relatively narrow gaps to the right and below the goal region. These trajectories, with a more sophisticated and coherent quantification of risk, are generated with the same computational complexity as with Gaussian chance constraints. 


\section{Conclusion} \label{sec_conclusions}
In this paper, we presented a methodological framework aimed towards tighter integration of perception and planning in autonomous robotic systems. The environmental state is estimated from sensor data to propagate both estimates and uncertainties of both robot and obstacles. Risk constraints are posed in a meaningful and coherent manner through distributionally robust chance constraints. Using a dynamic output feedback controller together with the distributionally robust risk constraints, a new algorithm called \texttt{OFDR}-$\rrtstar$ is shown to produce risk bounded trajectories with coherent risk assessment. Future research involves studying distribution propagation for nonlinear systems with higher order moments and considering Kalman filter variations (e.g., unscented) along with more sophisticated steering methods to explicitly incorporate the nearby obstacle constraints. Also, the combination of moment- and data-based distribution parameterizations for uncertainty modeling could be used to combine their relative advantages.
\bibliography{ifacconf}             
\end{document}